%% file: main.tex
\documentclass[conference]{IEEEtran}
\usepackage[utf8]{inputenc}

\usepackage{epsfig}
\usepackage{graphicx}
\usepackage{epstopdf}
\usepackage{balance}
\usepackage{url}
\usepackage{subfigure}
\usepackage{eurosym}
\usepackage{upgreek}
\usepackage{tabularx}
\usepackage{threeparttable}
\usepackage{booktabs}
\usepackage{lipsum}
\usepackage{mathtools}
\usepackage{color}
\usepackage{xcolor}
\usepackage{cite}
\usepackage{amsmath,amssymb,amsfonts}
\usepackage{algorithmic}
\usepackage{textcomp}
\def\BibTeX{{\rm B\kern-.05em{\sc i\kern-.025em b}\kern-.08em
    T\kern-.1667em\lower.7ex\hbox{E}\kern-.125emX}}

\usepackage{geometry}
\begin{document}

%

\title{Towards a Wearable Interface for Food Quality Grading through ERP Analysis}

\vspace{-5mm}
\author{\IEEEauthorblockN{M. Guermandi\IEEEauthorrefmark{1},
S. Benatti\IEEEauthorrefmark{1},
D. Brunelli\IEEEauthorrefmark{2},
V. Kartsch\IEEEauthorrefmark{1}, 
L. Benini\IEEEauthorrefmark{1}\IEEEauthorrefmark{3}}
\IEEEauthorblockA{\IEEEauthorrefmark{1}DEI, University of Bologna, Italy. Email:  
\{marco.guermandi, simone.benatti, victorjavier.kartsch, luca.benini\}@unibo.it}
\IEEEauthorblockA{\IEEEauthorrefmark{2}DII, University of Trento. E-mail: davide.brunelli@unitn.it}
\IEEEauthorblockA{\IEEEauthorrefmark{3}Integrated System Laboratory, ETHZ, Zurich, Switzerland. Email: lbenini@iis.ee.ethz.ch}
\vspace{-0.2in}
}



%



\maketitle

\begin{abstract}
Sensory evaluation is used to assess the consumer acceptance of foods or other consumer products, so as to improve industrial processes and marketing strategies. The procedures currently involved are time-consuming because they require a statistical approach from measurements and feedback reports from a wide set of evaluators under a well-established measurement setup. In this paper, we propose to collect directly the signal of the perceived quality of the food from Event-related potentials (ERPs) that are the outcome of the processing of visual stimuli. This permits to narrow the number of evaluators since errors related to psychological factors are by-passed. We present the design of a wearable system for ERP measurement and we present preliminary results on the use of ERP to give a  quantitative measure to the appearance of a food product. The system is developed to be wearable and our experiments demonstrate that is possible to use it to identify and classify the grade of acceptance of the food.


\end{abstract}

\input{./content/01_intro.tex}
\input{./content/02_system.tex}
\input{./content/03_exp_results.tex}
\input{./content/04_conclusions.tex}

\section*{Acknowledgment}
This work has been partially supported by the European H2020 FET project
OPRECOMP under Grant 732631



\IEEEpeerreviewmaketitle

\bibliographystyle{IEEEtran}
\bibliography{IEEEabrv,BIOCAS16}

\end{document}

%% file: content/01_intro.tex
\section{Introduction}
In food industry, sensorial analysis is one of the tools that industrial process and marketing management can use to understand the target market and to optimize the effort and investment during product development. Sensory evaluation is defined as the scientific method used to give a quantitative measure to the appearance or flavour of a food product as perceived through the senses of sight, taste, smell, touch, and hearing \cite{ STONE19931}.
It is not just a collection of "likes and dislikes“  but this discipline scientifically determines, measures, and interprets physiological responses to physical stimuli produced by a food product.
The sensorial analysis is becoming an irreplaceable tool in food industry for planning the success of a product. In fact, when consumers buy food products, they are mostly driven by personal sensorial feedback and by the brand, rather than other important features like nutrition elements or convenience.

Sensorial analysis methods can be classified into two classes: affective, or hedonic, and analytical methods \cite{Hofer2013}. Affective methods use large consumer panels, or several trained panelists, to answer a long questioner after having tasted the products, following a well-defined procedure.  Hedonic methods require a large panel size, because results can be interpreted only with statistical tools to have high confidence in the interpretation of the consumers feedback. 
On the other side, analytical methods are specific tests done done by trained experts. Analytical methods can be used only to determine if products are different, or if a food variety highlights a selected characteristic more than another. Such methods follow a standardized ISO methodology \cite{ISO}. 
Psychological and sensorial feedback from stimuli produced by food products play a fundamental role also in ordinary industrial transformation of food (e.g. fruit and vegetables) as quality assessment during the whole transformation chain, usually done by a manual screening of the operators.

Therefore, sensory evaluation is an integral part of the food industrial process and of the success of a product. Nevertheless, two major factors limit this method. 
Firstly, it is time-consuming. When used for market analysis, forming large panels of evaluators and training them for a good evaluation campaign is expensive and takes several weeks to set up. In the ordinary industrial process, the manual screening from operators is time-consuming, tedious, labor-intensive and expensive, and rarely applied at bulk trading points \cite{bhatt2015automatic}. 
The second limitation consists in the possible error of the human activity in the loop. All the adopted methods are prone to errors, driven by psychological factors, that may be introduced even by the most expert judge. For example, the evaluator may give undifferentiated scores to the samples in the central part of the scale, for fear of making mistakes; or may consider two different attributes very similar (e.g. “ripe” and “juicy”) and thus rating them in a similar manner. Finally, such methods may be hampered by maliciousness or capriciousness of the evaluator that can misrepresent the actual physiologic sensorial perception \cite{Sensory2010}.

In this paper, we propose to collect directly the signal of the perceived quality of the food from Event-related potentials (ERPs) that are the outcome of the processing of visual stimuli, and to process them in order to have an objective evaluation of the sensory perception, without any bias introduced by the operator through higher brain functions.

ERPs are one of the most common technique to study emotion processing by visual stimuli \cite{schupp2000affective}. Evidence shows how this response is modulated by motivational relevance of the presented pictures. This has been applied to several fields, including the study of craving towards food, nicotine and recreational drugs \cite{littel2007effects}. Other studies \cite{venkatraman2015predicting}, have proposed to complement traditional measures with physiological measures to improve the advertising models. In this work, we start to adopt this kind of measures to assess the perceived quality of food pictures. Then, these metrics can be used to train an automated grading system based on machine learning from visual inspection. In particular, we focus in acquiring and decoding brain activity response to food pictures.

In the last twenty years, several works \cite{schupp2000affective} have shown how ERPs (the averaged brain response recorded with scalp electrodes to a recurrent stimuli, e.g. a visual one) vary with the judged emotionality of visual stimuli. Specifically, a late positive  potential (LPP) is enhanced for stimuli evaluated as distant from an established context. A pleasant stimulus presented within a series of  unpleasant pictures elicits a larger LPP than does the same pleasant target, presented among other pleasant stimuli (so called \emph{oddball paradigm}). A different paradigm has also been used in which pleasant, neutral, and/or unpleasant pictures appear with equal probability in a random sequence. Substantial late positive shifts in the ERPs are evoked by emotional pictures both  pleasant and unpleasant, compared with neutral images, similarly over left and right lateral recording sites. This shifts can begin as early as 200 ms after onset and last over a long period of picture presentation.

We present a wearable system for the detection of ERPs with quick setup features which can be adopted outside the laboratory setting and we demonstrate that it is able to detect different responses to two different food grades while also embedding the required signal processing. In this preliminary study, we distinguish between commercial and non-commercial grade apples, but we plan to extend the analysis to different types of food and finer-granularity grading scale. We also show that, while running the application, the device consumes only 10.88 mW, providing up to 19 hs of operation with a battery life that is further extended up to 40 hs when energy is generated through an Energy Harvesting (EH) subsystem. The full software runs on the wearable node that employs less than 10\% of the total power to process the data. Thus, the remaining power can be employed on power-demanding high-quality AFEs, resulting in an improvement of the overall performance of the system. 

The system is based on BioWolf, an integrated platform for computationally-intensive medical IoT applications, which features the capability to acquire biosignals (up to 8 channels, 32 ksps, noise level compatible with high quality EEG acquisition) and to process them locally on an ULP computing platform with a power budget lower than that of the analog front end. Our platform is based on Mr. Wolf \cite{Wolf}, a programmable Parallel Ultra-Low-Power processor that combines high versatility and computational efficiency higher than single-core architectures such as those available in standard MCUs with wireless connectivity. 

Details of the system design are described in Section \ref{sec2}, both from the hardware and data processing standpoints. Section \ref{sec3} presents a characterization of the system, specifically in terms of power consumption and capability to distinguish brain activity response to the two different class of pictures. In Section \ref{sec4} we draw some conclusions and outline future work.

%% file: content/02_system.tex
\section{System Description}\label{sec2}
\begin{figure}[t]
\begin{center}
\includegraphics[width=0.75\columnwidth]{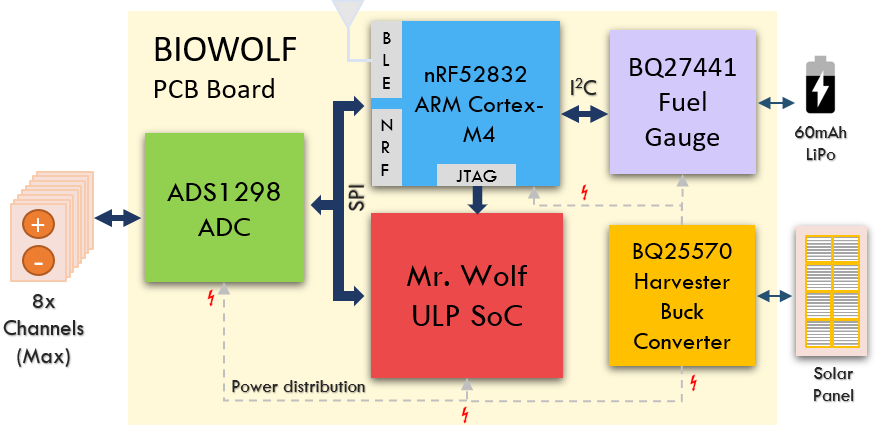}
\caption{BioWolf System Architecture.}
\label{fig:sys_archi_blocks}
\vspace{-5mm}
\end{center}
\end{figure}

\begin{figure}[t]
\begin{center}
\includegraphics[width=0.65\columnwidth]{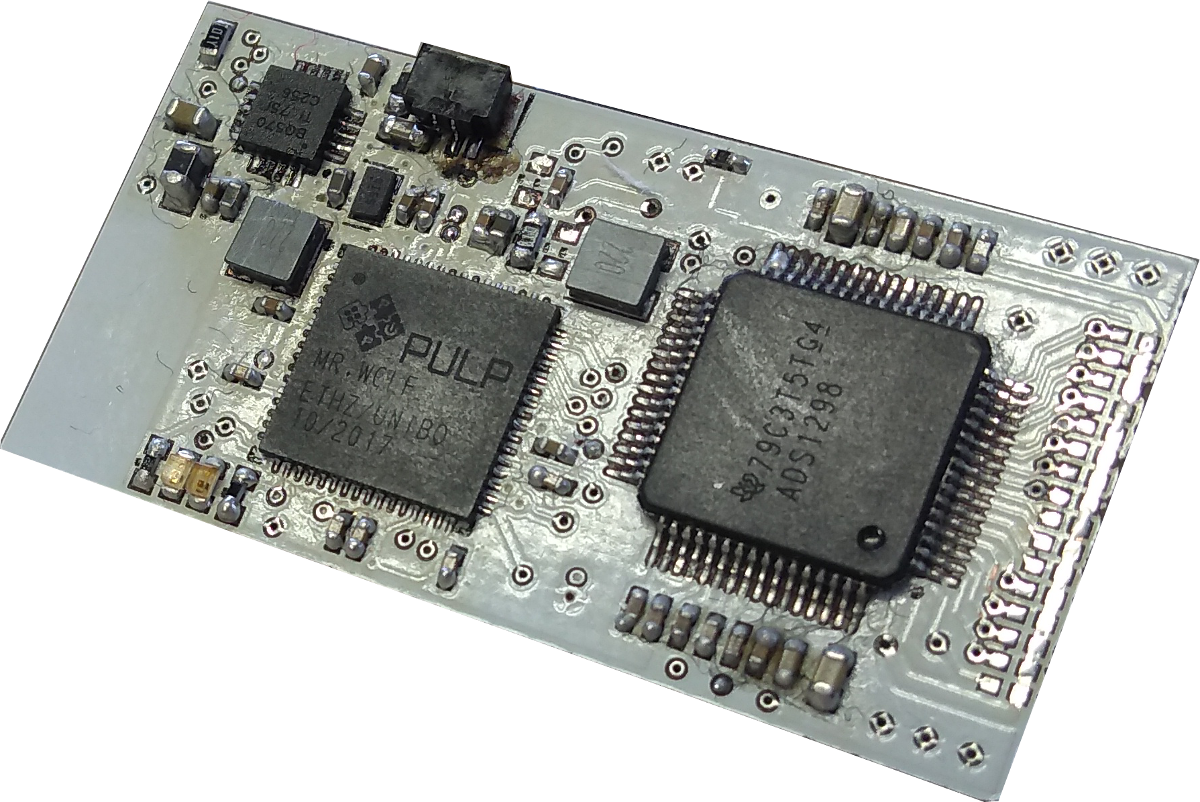}
\caption{BioWolf Board. Top side allocates Mr. Wolf, the AFE and part of the power supply section. Bottom side is mostly dedicated to the nRF52832 SoC, fuel gauge, connectors and the analog power supply section.}
\label{fig:sys_archi}
\vspace{-5mm}
\end{center}
\end{figure}

BioWolf is a highly-configurable platform for acquisition and embedded processing of biopotentials. It features a Parallel Ultra-Low-Power (PULP) SoC MCU for embedded signal processing, an ARM-based SoC for Bluetooth Low Energy (BLE) communication and system control and management, and an Analog Front End (AFE) for analog-to-digital conversion of up to 8 differential biosignal channels. It also includes a nano-power buck-boost regulator for energy harvesting (e.g. from solar source) and a  fuel gauge to check for battery status. Fig. \ref{fig:sys_archi_blocks} shows a block diagram of the complete system and Fig. \ref{fig:sys_archi} shows the final PCB implementation.
%

BioWolf can operate in three different modalities to provide maximum flexibility while minimizing power consumption.  
\begin{itemize}
\item When on-board processing is not required or is very limited (such as for basic filtering) and data needs to be streamed out directly, Mr. Wolf is put in sleep mode and the Nordic SoC acts as master on the SPI bus, reading data directy from the AFE and streaming them out to the host via BLE. 
\item When more computationally intensive processing is required, Mr. Wolf guarantees the best power efficiency to the system. The ULP processor directly controls the SPI bus as the master, while the Nordic SoC and the AFE are both slave peripherals. Data is read from the AFE, processed and only the result of such processing is sent to the Nordic SoC and from there to the host through BLE.
\item A deep sleep mode is available to minimize power consumption when the system is not required to acquire and/or process data. The device can be woken up by exposing it to an NFC field, such as tapping on it with a NFC-enabled smart-phone or tablet.
\end{itemize}
 
Interface with subject skin is obtained through commercial dry-electrode contacts (g.SAHARA from g.tec)\cite{guger2012comparison}, so as to minimize setup time and reduce discomfort with respect to standard wet electrodes requiring skin preparation. Custom active electrodes are used to make signal quality resilient to the higher contact impedance of dry electrodes with respect to wet electrodes with skin preparation. As single-ended amplification stages with gain higher than one limit the rejection of common mode noise \cite{salvaro2018minimally}, only signal buffering is performed on the active electrode by an Operational Amplifier (O.A.) connected as a unity-gain buffer. A 68 K$\Omega$ resistor in series with the amplifier input limits the patient auxiliary current below the applicable limit of 50 $\mu$A in single fault condition (i.e. electrode shorted to one of the power supplies). The O.A. is an AD8603 from Analog Devices, showing low voltage noise (2.3 $\mu$V peak-to-peak in the 0.1 to 10 Hz band and 25 nV/$sqrt(Hz)$ at 1 KHz) and a maximum quiescent current of 50 $\mu A$. Input impedance is in excess of 500 M$\Omega$ in the EEG band.
 
The outputs of the active electrodes are acquired by a multichannel commercial AFE from TI (ADS1298), which can be configured to acquire up to 8 channels at sampling rates up to 32 ksps. Input can be configured in differential and single-ended mode, with a gain of the input programmable gain amplifier (PGA) from 1 to 12 and a maximum resolution of 24-bits. The system supports active electrodes with both dry- and wet-contact.
The ADSXXXX family of AFEs is the \textit{de-facto} standard for biopotential acquisition as they present a very favorable trade-off between performance and power consumption and between size and number of channels. With respect to ADS1299 which targets EEG signal acquisition, the adopted ADS1298 has reduced power consumption and allows for 2.7 V supply operation (removing the need for step-up DC/DC conversion of the battery voltage) without significantly degrading noise performance \cite{guermandi2018wearable}. 

Mr. Wolf is a multi-core programmable SoC implemented in CMOS 40nm technology that combines a tiny (12 Kgates) RISC-V processor (zero-risky) \cite{RISCV}, namely the Fabric Controller (FC), with a cluster of eight RISC-V processors equipped with flexible and powerful DSP extensions available on the RI5CY processor \cite{RISCV}. The cluster is coupled with a single-cycle latency multi-banked L1 memory (64 kB) allowing fast data transfer among the cores, and with an 'off the cluster' 512 kB of memory (L2) with 15 cycles latency. A dedicated DMA controller allows reducing the latency and computational power associated with data transfer. It also features two floating-point units (FPU) that are shared among the cores. Mr. Wolf can achieve very fine-grained parallelism and high energy efficiency in parallel workloads through a dedicated hardware block (HW Sync) that provides fast event management, parallel thread dispatching and synchronization. The SoC contains a full set of peripherals, including a Quad SPI (QSPI), I2C and UART, with data transfers also managed by a multi-channel I/O DMA to reduce the load on the system. In run mode, the SoC is powered by an internal DC/DC converter that can be programmed to deliver from 0.8 V to 1.1 V. In sleep mode, a low-dropout (LDO) regulator powers a real-time clock (32 kHz crystal oscillator) that controls a programmed wake-up and, optionally, part of the L2 memory, allowing retention of application state for fast wake-up. In deep sleep mode, the power consumption of the MCU is about 108 $\mu$W that can be further reduced to 72 $\mu$W when no retention is required.

Data communication with a remote host (and eventually basic processing if needed) is performed by the nRF52832 SoC from Nordic. The MCU is based on an ARM Cortex-M4 core running up to 64 MHz clock frequency and provides flexible Bluetooth 5 (BLE) communication at a low-power budget. This MCU also serves as a device manager for the board. It allows choosing the operation mode (sleep, raw data streaming, data acquisition and processing) and checking battery charging operation and status.
As Mr. Wolf does not embed any non-volatile memory, its firmware is stored in a portion of the Nordic SoC Flash memory. Programming is performed by the Nordic SoC itself by transferring the code to Mr. Wolf L2 memory through JTAG interface.

Power supply, battery management, and EH from solar sources are managed by a Texas Instruments BQ25570. The IC implements a Maximum Power Point Tracking (MPPT) that adapts the input impedance of the solar cells maximizing the energy conversion in all the lighting conditions with up to 90\% of efficiency. This energy is then used to recharge a small form factor 60 mAh LiPo battery. The EH also provides a high-efficiency buck converter that delivers a stable voltage output of 1.8 V to supply the digital portions of the board. An additional output is available, connected to the battery voltage when its voltage level is higher than 3 V. This is used to power the analog portions of the board, in particular, the analog circuitry of the AFE which is powered at its minimum supply voltage of 2.7 V by a low noise linear regulator.

\subsection{System Setup}
\begin{figure}[t]
\begin{center}
\includegraphics[width=0.95\columnwidth]{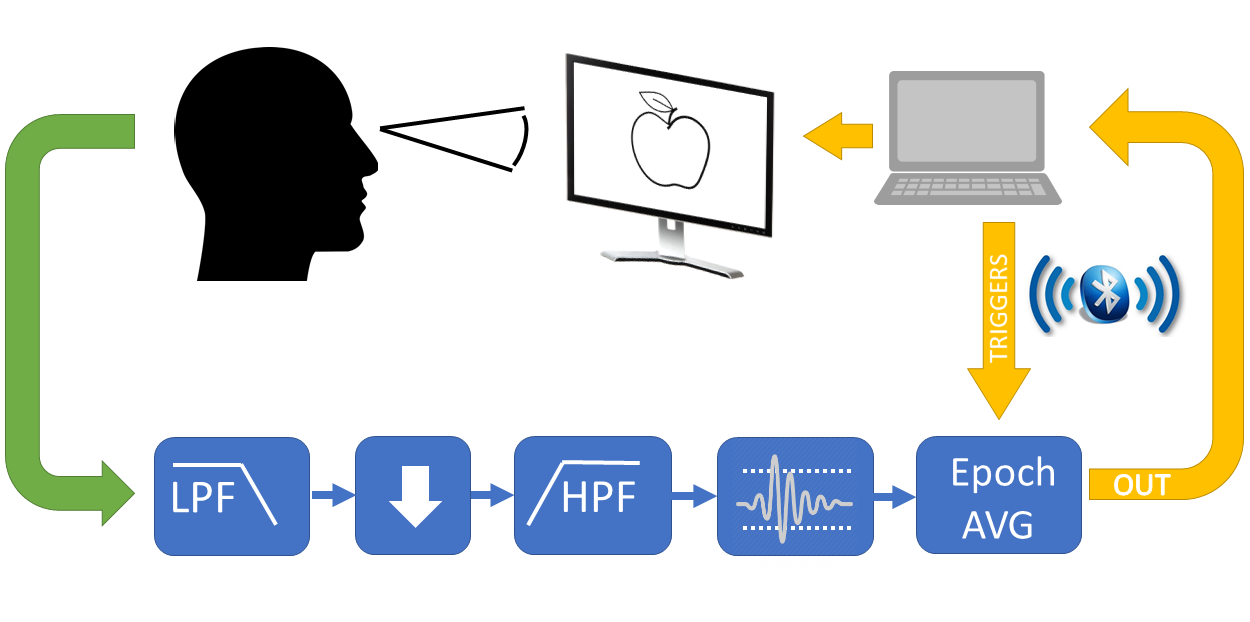}
\caption{Setup and processing steps of the presented system. Images are presented on a 17 inch LCD screen, while EEG is recorded in PO7 and PO8 with reference on Fz. Data is band-pass filtered between 0.25 and 30 Hz and decimated by a factor 5 to 100 sps. Epochs with data above a $\pm$ 50 $\mu$V are discarded. The remaining epochs are averaged to provide the final EPRs.}
\label{fig:processing}
\vspace{-7mm}
\end{center}
\end{figure}

To minimize setup time and improve user friendliness, the EEG signal is acquired on three electrodes only. Electrode positions (Fz, PO7 and PO8 according to the 10/20 system) have been  selected through preliminary testing. The electrodes in the parietal-occipital region capture the visual processing of the image, while the signal measured in the frontal region reflects the cognitive processing of the image. They can be acquired at the same time with a setup as simple as a headband.
Images are presented on a 17-inches LCD screen from a PC running Psychtoolbox under MATLAB. Setup and processing steps are presented in Fig.\ref{fig:processing}.

\subsection{Data Processing}
Data processing is performed on Mr. Wolf. Since computation is relatively simple, all processing can be performed on the FC which  processes data in double-precision fixed point representation. It also takes care of exchanging data with the AFE for signal acquisition and with the nRF32832 SoC for communication with the host through BLE. The latter is a bidirectional communication as the host needs to provide synchronization signals (\emph{triggers}) to inform the device of the instant at which the visual stimulus is presented to the subject. When acquisition of a complete trial is completed, the device sends the reconstructed ERPs to the host.\\
\\
The processing steps for reconstruction of the ERPs are the following:

\begin{itemize}
\item \emph{Preprocessing:} the first operation is the computation of the average of data acquired from the two differential channels (PO7-Fpz and PO8-Fpz) at 500 SPS. Data is then filtered with a Finite Impulse Response (FIR) low-pass filter (LPF), with 100 taps and -3dB corner frequency of 30 Hz. This allows to down-sample data of a factor 5, reducing the computational burden of the filtering steps as both the outputs of the LPF and the subsequent High-Pass Filter (HPF) can be computed at a reduced sampling frequency of 100 SPS. FIR filtering is preferred to IIR despite the increased computational burden because of its linear phase which allows to minimize distortion in the reconstructed ERP. Since we are interested in late potentials in the ERP, high pass filtering needs to be performed at very low corner frequency of 0.25 Hz. This requires the use of an extremely high order filter (1000 taps) which dominates the computational burden of the ERP reconstruction algorithm.
\item \emph{Epochs reconstruction:} to remove epochs containing artifacts we adopted a simple method based on automatically rejecting epochs containing samples over $\pm 50 \mu V$. To this purpose, each sample at the output of the HPF is checked at run-time and, if its absolute value is higher than the threshold, the epoch is rejected and every computation for that epoch (including filtering of subsequent samples) is stopped. Information on the time at which the epoch starts is sent from the host via BLE. When the epoch is not rejected, it is averaged with previously accepted epochs to reconstruct the ERP.

\end{itemize}

%% file: content/03_exp_results.tex
\section{Experimental results}\label{sec3}
\subsection{Power Consumption}
To evaluate the performance of the system, we set the operating frequency and voltage of Mr. Wolf at 50 MHz and 0.8 V, respectively. Although this frequency can be scaled up to 450 MHz to meet more constrained applications, the processing required in this work is minimal. Thus, a lower operational frequency satisfies the real-time constraints while minimizing the overall power consumption.

The power consumption of the system is the contribution of the active blocks, namely, Mr. Wolf, the ADC, and the Nordic SoC, for a total of 10.88 mW. Data sampling through the analog sections (ADS1298) requires the highest share of the overall power (around 80\%). Although the use of this ADC comes at a high power cost, it ensures the required signal quality, also allowing to avoid more rigorous filtering of the signal that could be translated into higher power consumption at the MCU side.
The digital section that includes occasional BLE transmission of computation results and synchronization of the trigger, and the data transfers between AFE and Mr. Wolf, represents the 11\% of the total power. 

Mr. Wolf is responsible for the remaining power, which is the result of power management techniques, such as switching thought the MCU power modes. While working in Run Mode, the acquisition and processing of a single sample only requires 0.5 ms and, when in idle, the MCU is put into sleep mode to minimize the power consumption. Since the FC is able to satisfy the real-time requirements of the application, the cluster remains off all the times. This demonstrates the versatility of Mr. Wolf to work at a low power budget throughout different computational needs. Future work that might include heaver DSP will take advantage of the cluster, able provide heavier computing of the kernel functions. 

As a result, our system achieves up to 19 h of autonomy with a 60 mAh battery, which can be further extended up to 20 h and 40 h in indoor (600 lux)/outdoor (10000 lux) scenarios, respectively, using the EH subsystem. 

\subsection{ERP Analysis}
\begin{figure}[t]
\vspace{-4mm}
\centering
\includegraphics[width=0.7\columnwidth]{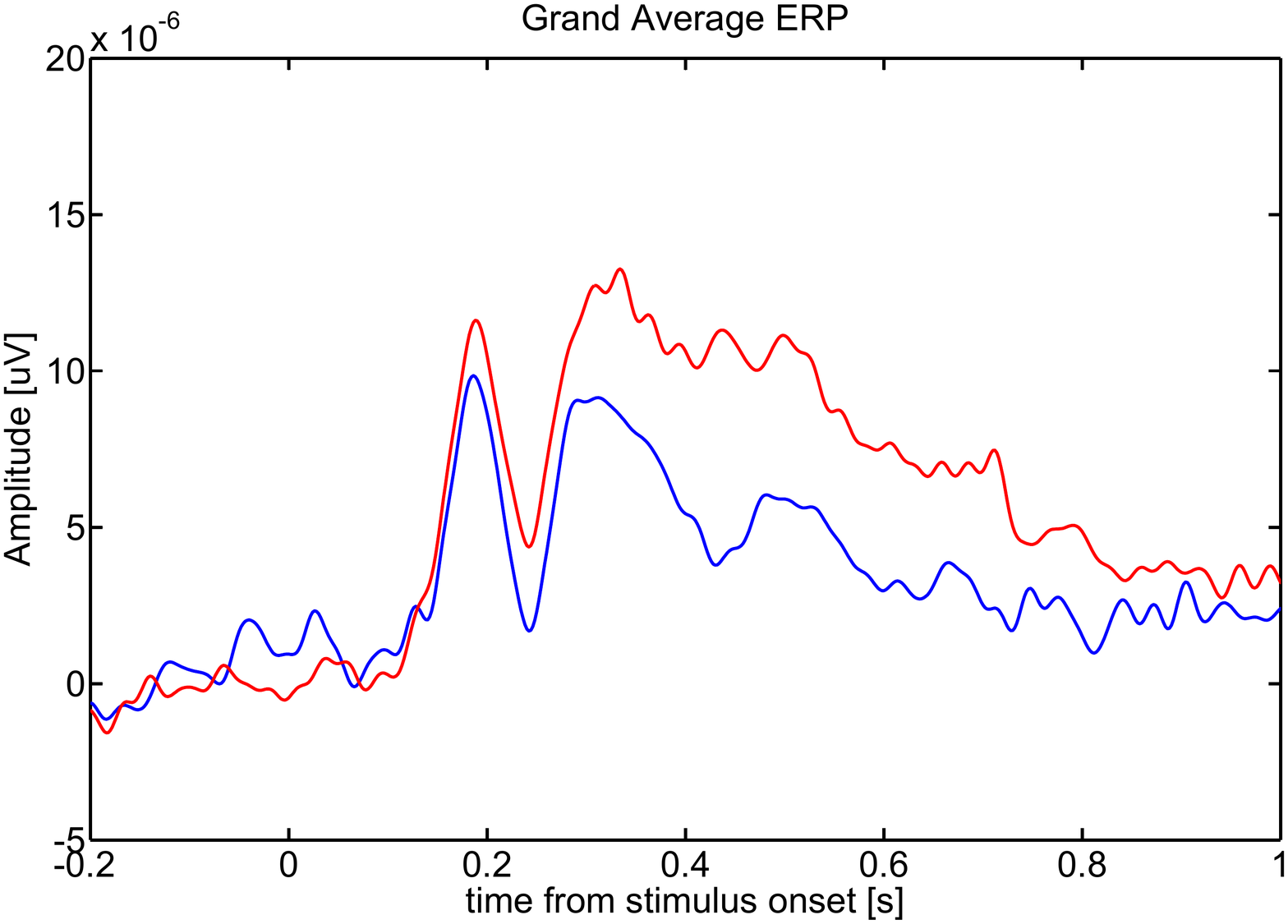}
\vspace{-2mm}
\caption{Grand average of the ERPs extracted from the five subjects. Red line corresponds to non-commercial grade apple pictures, blue line to commercial-grade ones. Data is band-pass filtered between 0.25 and 40 Hz, epochs are rejected if EEG amplitude exceeds $\pm$ 50 uV. \label{overflow}}
\label{fig:ERP}
\vspace{-3mm}
\end{figure}

We tested the device on five able-bodied, (aged 26-42) without previous history of neurological disorders. All participants provided written consent to take part to the experiments. The subject was sit in a dimly lighted room, approximately 80 cm apart from a 17-inches LCD screen.
200 images of good-quality and defective fruits (100 per class) were presented randomly on a 17-inches LCD monitor for 1 second, separated by a 1 second fixation cross. Event Related Potentials were measured as the average potential between electrodes PO7 and PO8 versus Fz. Fig. \ref{fig:ERP} presents results obtained through the analysis of ERPs on the five subjects (grand average on top) and on a single subject (bottom). 

Averaged responses are flat before presentation of the stimulus (t=0) and present a high correlation till approximately 300 ms (ERP components related to visual processing of the image). After that time, late positive potentials (LPP) components show significant differences in response to commercial grade vs. non-commercial grade fruit images, testifying how responses to the two classes of stimuli are distinguishable, and giving a clear indication on the suitability of this approach for automated quality grading analysis.

\begin{figure}[t]
\vspace{-4mm}
\centering
\includegraphics[width=0.7\columnwidth]{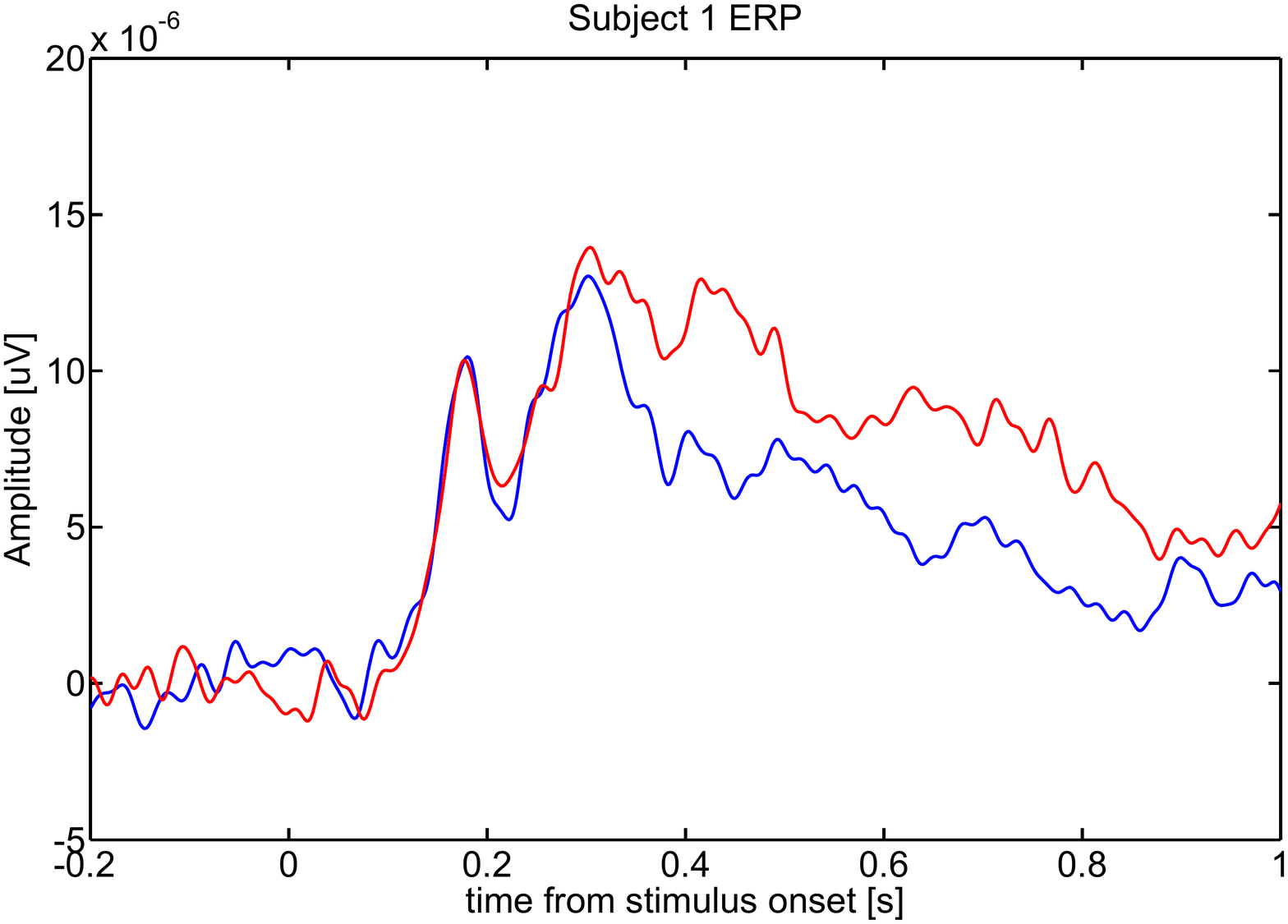}
\vspace{-2mm}
\caption{ERP from subject 1. Red line corresponds to non-commercial grade apple pictures, blue line to commercial-grade ones. Data is band-pass filtered between 0.25 and 40 Hz, epochs are rejected if EEG amplitude exceeds $\pm$ 50 uV. }
\vspace{-3mm}
\end{figure}

%% file: content/04_conclusions.tex
\section{Conclusions}\label{sec4}

This work presented an ERP measurement system featuring BioWolf, a Parallel Ultra Low Power platform, which allows EEG signal acquisition and processing that allow on-board, online and real-time differentiation of classes of images, with the objective of ranking food quality. In the presented work we show that our system detects 2 distinct ERPs responses to commercial and non-commercial grade apples. The processing of the EEG signal is performed on-line on the Mr. Wolf platform, and, by virtue of its energy efficiency, the system exceeds 19h of battery life with a tiny 60 mAh LiPo battery. Future work will target a finer grain classification of ERPs for visual stimuli presentation and the exploration of more advanced optimization and classification strategies.

%% file: main.bbl
\begin{thebibliography}{10}
\providecommand{\url}[1]{#1}
\csname url@samestyle\endcsname
\providecommand{\newblock}{\relax}
\providecommand{\bibinfo}[2]{#2}
\providecommand{\BIBentrySTDinterwordspacing}{\spaceskip=0pt\relax}
\providecommand{\BIBentryALTinterwordstretchfactor}{4}
\providecommand{\BIBentryALTinterwordspacing}{\spaceskip=\fontdimen2\font plus
\BIBentryALTinterwordstretchfactor\fontdimen3\font minus
  \fontdimen4\font\relax}
\providecommand{\BIBforeignlanguage}[2]{{%
\expandafter\ifx\csname l@#1\endcsname\relax
\typeout{** WARNING: IEEEtran.bst: No hyphenation pattern has been}%
\typeout{** loaded for the language `#1'. Using the pattern for}%
\typeout{** the default language instead.}%
\else
\language=\csname l@#1\endcsname
\fi
#2}}
\providecommand{\BIBdecl}{\relax}
\BIBdecl

\bibitem{STONE19931}
H.~Stone and J.~L. Sidel, ``Sensory evaluation practices (second edition),''
  ser. Food Science and Technology, H.~Stone and J.~L. Sidel, Eds.\hskip 1em
  plus 0.5em minus 0.4em\relax London: Academic Press, 1993, pp. 1 -- 17.

\bibitem{Hofer2013}
M.~von Meyer-Höfer, V.~von~der Wense, and A.~Spiller, ``Characterising
  convinced sustainable food consumers,'' \emph{British Food Journal}, vol.
  117, pp. 1082--1104, 03 2015.

\bibitem{ISO}
{International Organization for Standardization ISO}, ``Sensory analysis
  methodology. general guidance for establishing a sensory profile,'' ISO
  13299:2016.

\bibitem{bhatt2015automatic}
A.~K. Bhatt and D.~Pant, ``Automatic apple grading model development based on
  back propagation neural network and machine vision, and its performance
  evaluation,'' \emph{Ai \& Society}, vol.~30, no.~1, pp. 45--56, 2015.

\bibitem{Sensory2010}
T.~N{\ae}s, P.~B. Brockhoff, and O.~Tomic, \emph{Statistics for Sensory and
  Consumer Science}.\hskip 1em plus 0.5em minus 0.4em\relax Wiley, 2010.

\bibitem{schupp2000affective}
H.~Schupp \emph{et~al.}, ``Affective picture processing: the late positive
  potential is modulated by motivational relevance,'' \emph{Psychophysiology},
  2000.

\bibitem{littel2007effects}
M.~Littel and I.~H. Franken, ``The effects of prolonged abstinence on the
  processing of smoking cues: an erp study among smokers, ex-smokers and
  never-smokers,'' \emph{Journal of Psychopharmacology}, vol.~21, no.~8, pp.
  873--882, 2007.

\bibitem{venkatraman2015predicting}
V.~Venkatraman \emph{et~al.}, ``Predicting advertising success beyond
  traditional measures: New insights from neurophysiological methods and market
  response modeling,'' \emph{Journal of Marketing Research}, 2015.

\bibitem{Wolf}
A.~Pullini, D.~Rossi, I.~Loi, A.~D. Mauro, and L.~Benini, ``Mr. wolf: A 1
  gflop/s energy-proportional parallel ultra low power soc for iot edge
  processing,'' in \emph{ESSCIRC 2018 - IEEE 44th European Solid State Circuits
  Conference (ESSCIRC)}, Sept 2018, pp. 274--277.

\bibitem{guger2012comparison}
C.~Guger, G.~Krausz, B.~Z. Allison, and G.~Edlinger, ``Comparison of dry and
  gel based electrodes for p300 brain--computer interfaces,'' \emph{Frontiers
  in neuroscience}, vol.~6, p.~60, 2012.

\bibitem{salvaro2018minimally}
M.~Salvaro, S.~Benatti, V.~Kartsch, M.~Guermandi, and L.~Benini, ``A minimally
  invasive low-power platform for real-time brain computer interaction based on
  canonical correlation analysis,'' \emph{IEEE Internet of Things Journal},
  2018.

\bibitem{guermandi2018wearable}
M.~Guermandi, S.~Benatti, V.~J.~K. Morinigo, and L.~Bertini, ``A wearable
  device for minimally-invasive behind-the-ear eeg and evoked potentials,'' in
  \emph{2018 IEEE Biomedical Circuits and Systems Conference (BioCAS)}.\hskip
  1em plus 0.5em minus 0.4em\relax IEEE, 2018, pp. 1--4.

\bibitem{RISCV}
P.~D. Schiavone \emph{et~al.}, ``Slow and steady wins the race? a comparison of
  ultra-low-power risc-v cores for internet-of-things applications,'' in
  \emph{PATMOS}, Sept 2017, pp. 1--8.

\end{thebibliography}
